\documentclass[journal=jacsat,manuscript=article]{achemso}

\usepackage{chemformula} % Formula subscripts using \ch{}
\usepackage[T1]{fontenc} % Use modern font encodings
\usepackage{multirow}
\usepackage{lineno}

\author{Ming Geng}
\affiliation{Science Institute, University of Iceland, VR-III, Rekjav\'{i}k 107, Iceland}
\email{gengm@hi.is}
\author{Hannes J\'{o}nsson}
\affiliation{Science Institute, University of Iceland, VR-III, Rekjav\'{i}k 107, Iceland}
\alsoaffiliation{COMP Centre of Excellence and Department of Applied Physics, Aalto University, FIN-00076 Espoo, Finland}
\email{hj@hi.is}

\title{Site preference of Fe atoms in the olivine $(Fe_xMg_{2-x})SiO_4$ and its surface}

\begin{document}

%%%%%%%%%%%%%%%%%%%%%%%%%%%%%%%%%%%%%%%%%%%%%%%%%%%%%%%%%%%%%%%%%%%%%
%% The "tocentry" environment can be used to create an entry for the
%% graphical table of contents. It is given here as some journals
%% require that it is printed as part of the abstract page. It will
%% be automatically moved as appropriate.
%%%%%%%%%%%%%%%%%%%%%%%%%%%%%%%%%%%%%%%%%%%%%%%%%%%%%%%%%%%%%%%%%%%%%
%\begin{tocentry}

%\end{tocentry}
%%%%%%%%%%%%%%%%%%%%%%%%%%%%%%%%%%%%%%%%%%%%%%%%%%%%%%%%%%%%%%%%%%%%%
%% The abstract environment will automatically gobble the contents
%% if an abstract is not used by the target journal.
%%%%%%%%%%%%%%%%%%%%%%%%%%%%%%%%%%%%%%%%%%%%%%%%%%%%%%%%%%%%%%%%%%%%%
\begin{abstract}
	Olivine is involved in many natural reactions and industrial reactions as a catalyst. The catalytic ability is highly possible rely on the $Fe^{2+}$ in olivine. We use density functional theory calculation and thermodynamics to investigate the site preference of Fe atom in olivine which composition from iron-rich to iron-poor and its surfaces. The $Fe^{2+}$ always shows its high spin (quintet) state which has larger ion radius than $Mg^{2+}$ in olivine crystal and surfaces. The $Fe^{2+}$ inside the surface slab prefers the smaller M1 site than M2 site by enlarge the metal-oxygen octahedra when occupied the metal site as in the bulk system. Energy contribution of entropies accumulation caused temperature raise stops this preference at the temperature where a cation order-disorder distribution energy crossover happen in olivine. Surface exposed site provide $Fe^{2+}$ large space due its unsaturated nature. This lead a higher level of preference of $Fe^{2+}$ to the surface site than any metal site inside the crystal no matter M1 or M2 site is exposed. This indicate the $Fe^{2+}$  in the bulk system can diffuse to a metal site exposed on the surface driven by the energy difference. Many reactions can use the on surface $Fe^{2+}$ as a catalyst because of the active chemical behavior of Fe. Meanwhile this energetics preference should be considered in the future model to explain the natural observed zoning olivine have a high Fe edge and low Fe center.  These microscopic understanding can be essential to many olivine related geochemical and astrochemical reactions.
\end{abstract}

%%%%%%%%%%%%%%%%%%%%%%%%%%%%%%%%%%%%%%%%%%%%%%%%%%%%%%%%%%%%%%%%%%%%%
%% Start the main part of the manuscript here.
%%%%%%%%%%%%%%%%%%%%%%%%%%%%%%%%%%%%%%%%%%%%%%%%%%%%%%%%%%%%%%%%%%%%%
\section{I. Introduction}
\begin{figure*}[h]
	\centering
	\includegraphics[width=0.9\linewidth]{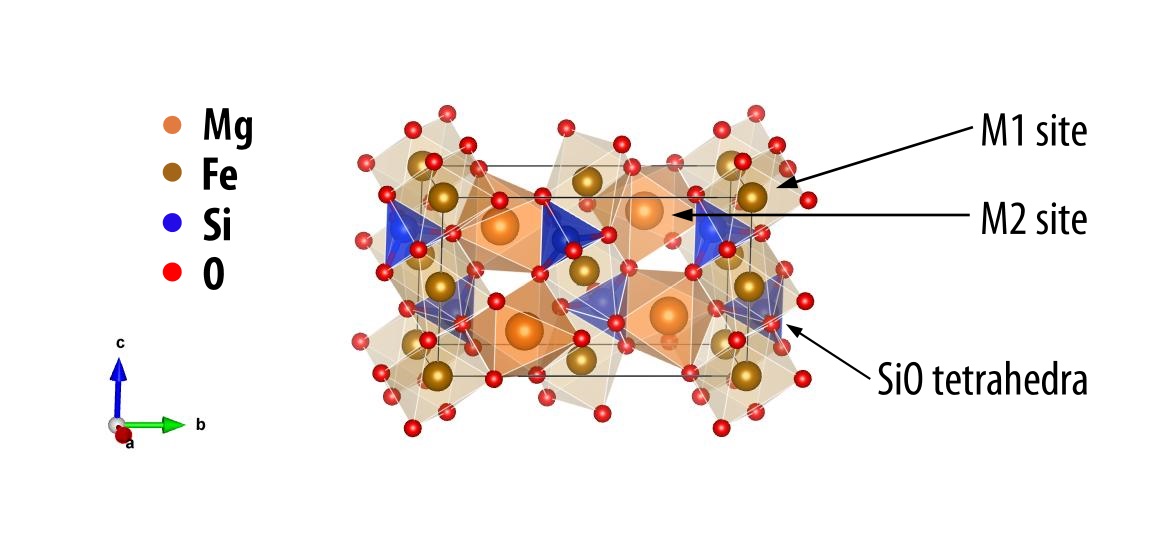}
	\caption{Structure of olivines crystal (Fo50 as an example) M1 and M2 site are two type of metal cation site, both of these sites with the oxygen around formed two types of octahedra. Silicon and oxygen atoms formed tetrahedra. }
	\label{fig:BulkStruct}
\end{figure*}

Olivine, a magnesium-iron silicate mineral $(Mg_xFe_{2-x})SiO_4$ ($x=0 - 2$), is the predominant mineral in both the Earth's upper mantle and interstellar media in space. Consequently, knowledge of the physical and chemical properties of olivine is of great geophysical and astrochemical interest because of its role in many important processes\cite{RN1443,RN1934,RN1942,RN1453,RN1447}. Olivine forms a solid solution series between the fayalite, iron end member ($Fe_2SiO_4$), and forsterite, the magnesium end member ($Mg_2SiO_4$). Fo value (forsterite percentage) usually used for describing the composition of the mineral in the solution, such as Fo50 means the olivine with 50\% of forsterite and 50\% fayalite. Olivine crystals have an orthorhombic structure with the space group $Pbmn$ each Si atom in olivine coordinates with four O atoms to form a $[SiO_4]$ tetrahedra, while each (Mg, Fe) atom are surrounded by 6 O atoms in two types of inequivalent metal site, one on the plane of mirror symmetry (M2 site) and the other on the inversion center (M1 site) as shown in Figure.\ref{fig:BulkStruct}. Forsterite transforms to spinel structure minerals, wadsleyite ($Imma$) and ringwoodite ($Fd3m$), in mantle transition zone with pressure upon to ~13.5GPa and ~18GPa \cite{RN1976}, respectively. Fayalite directly transforms to spinel structure at ~8GPa\cite{RN1985}. The natural olivine lays on a mixer of the Fe-Mg solid solution would expect to transform to spinel structure between 13.5GPa and 8GPa corresponding to the iron concentration based on thermodynamics mixing laws. Even with the same iron concentration, the distribution of Fe and Mg atoms also affects the physical and chemical properties of the mineral\cite{RN1895}. Since $Fe^{2+}$ has a more complex electronic configuration than $Mg^{2+}$, thus olivine with iron may lead more interesting chemistry than pure magnesium forsterite towards the reactivity of the fluid molecules in touch with a surface of the olivine if the $Fe^{2+}$ happen to expose on the surface which many catalysis studies have proved\cite{RN1921,RN1453}. On the other hand, the $3d^6$ in $Fe^{2+}$ also poses challenges due to the interplay of charge and spin at the Fe site. First thing need to know is the $Fe^{2+}$ distribution in the surface slab which is very difficult to investigate in experiment. Therefore combining the surface structure and $Fe^{2+}$ distribution in bulk is needed in further understanding these catalytic process. 

In some geophysical theoretic models, the $Fe^{2+}$ distribution over M1 and M2 sites were described using the value of a distribution coefficient($k_D$)
\begin{equation}
     k_D = \dfrac{Fe^{M1} \cdot Mg^{M2}}{Fe^{M2} \cdot Mg^{M1}} 
\end{equation}
which $Fe^{M1}$ and $Mg^{M1}$ are atomic fractions of $Fe^{2+}$ and $Mg^{2+}$ in M1 site, similar for M2 site. The $k_D$ larger than 1 is considered the Fe atom prefers more to the M1 site. Some mantle seismic discontinuities are considered to be related to the $k_D$ of olivine\cite{RN1899}. Observation found that olivines form the plutonic and metamorphic rock have a roughly equal $Fe^{2+}$ and $Mg^{2+}$ in both M1 and M2 sites\cite{RN1989}. Meanwhile, Moon's olivines and those from volcanic terrestrial rock seems to be a weak preference of $Fe^{2+}$ for M1 site. These observation indicate the different evolution process can lead different cation distribution when natural olivine crystallization. Studies in Li-ion batteries trying to utilise the multivalent ion rather than monovalent to improve the energy density of the battery. Since $(Mg,Fe)_2SiO_4$ has the same structure as the $LiFePO_4$, the $(MgFe)_2SiO_4$ also become a strong candidate of new battery cathode material. Orderer phase of $MgFeSiO_4$ structure have been synthesised through high temperature methods. However, the ordering distribution of $Mg$ and $Fe$ in the olivine which is the fundamental part to consider the possible ion diffusion mechanism is not clear yet.

In experimental studies, various results was conducted from different methods on the distribution of $Fe^{2+}$. It was found by many studies olivines shows a weak  ordering distribution of $Mg^{2+}$, $Fe^{2+}$ at high temperature and numerous quenching experiments.\cite{RN1793,RN1902,RN1986, RN1983}  Most experimental results were obtained by X-ray diffraction or M\"ossbauer spectroscopy and these methods are poorly applicable of to vast majority of natural olivines due to nearly all mantle olivines fall in Fo89 to Fo94\cite{RN1987} which have a relative low iron content. Two M\"ossbauer doublets caused by $Fe^{2+}$ in M1 and M2 site are very close to each other by both isometric shift and quadrupole splitting thus strongly overlapping in the spectra. New methods such as polarized near-infrared optical absorption spectroscopy \cite{RN1891} and paramagnetically shifted NMR \cite{RN1890} were tested to determine the site preference of the transition metal especially $Fe^{2+}$ in the olivine. Unfortunately,experimental results failed to reach a common conclusion. Atomic simulation of mineral crystal and surfaces become useful method to investigate the mineral in extreme pressure temperature conditions as in the Earth interior, surface structure and electron properties due to the difficulties in experiment researches. Most of theoretical studies are focus on the end members, some particular properties are investigated on certain iron concentration. Chartterjee et al.\cite{RN1172} compared $Fe^{2+}$ M1 and M2 in Fo50, whereas the natural olivines have a much higher Fo which means less iron content. A slight preference of M1 site of $Fe^{2+}$ were found in their calculation result. Das et al.\cite{RN1895} compared two type of cation ordering Fo50 olivine elastic properties. Javier et al.\cite{RN1449} is the only study compared pure magnesium forsterite and one iron containing olivine(Fo75) surface with B3LYP functional, but the cation site difference is not discussed in the study. A systematic analysis of the distribution of cations in olivine and its surface is needed to establish a ion diffusion mechanism and surface reactivity. A surface slab can be approximate as two exposed surfaces layer and a block of bulk system. In our work, we investigated the $Fe^{2+}$ distribution both in bulk and surface system in the same schema and combine them together, so we can get the $Fe^{2+}$ preference in a slab.

In this article, we present simulations of $(Mg_xFe_{2-x})SiO_4$ with several Fe concentrations in bulk and its surface to analyze the energy preference of the Fe atoms distribution in the material using Density functional theory (DFT) calculations. Section II will describe the crystal and surface structure and computation detail. In section III, the results obtained are presented with some discussion. Conclusion are summarized in section IV.

\section{II. Methods}

\subsection{Crystal and Surface structures}

\begin{figure*}[h]
	\centering
	\includegraphics[width=0.9\linewidth]{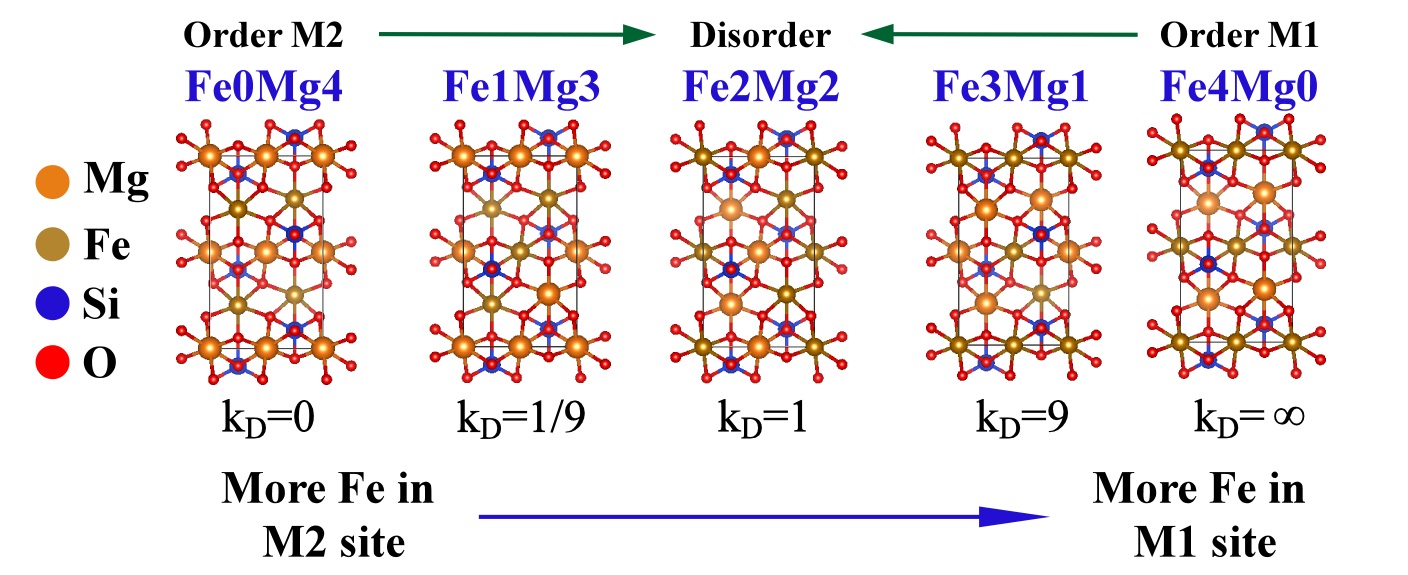}
	\caption{Side view of Fo50 structures (Mg:Fe = 4:4). The structure is named by the metal atom number in M1 sites. $k_D$ is the distribution coefficient. $k_D=1$ represent a fully mixed structure. The larger $k_D$ the more Fe atom distributed in the M1 site.  }
	\label{fig:Fo50Bulk}
\end{figure*}

Like many silicate mineral, olivine are essentially built out of $SiO_4$ tetrahedra blocks.The tetrahedra unites are isolated from each other and this type of mineral is called nesosilicates. The unit cell contains 4 formula units, 4 Si, 16 O atoms and 8 metal atoms. With the oxygen atoms on the tetrahedra corners, two types of octahedra also formed of which metal cations sites $M1$ and $M2$ are located in the center as shown in Figure.\ref{fig:BulkStruct}.The $M1O_6$ octahedra unites are connected to each other like a chain by sharing a $O-O$ edge formed by two oxygen different oxygen atoms. $M2O_6$ octahedra unites are connected by sharing two equivalence oxygen atoms to form a plane parallel to ac face of the unit cell. At room temperature, the $M1$ site is slightly more distorted than $M2$ \cite{RN1988}. So the M1 site usually thought to be preferred by smaller cation. Since olivine is a solid solution of magnesium and iron, we chose 15 bulk structures which the $Fe:Mg$ equal $2:6$, $4:4$ and $6:2$ in bulk unit cell which the iron contents are 25\%, 50\% and 75\% respectively to cover the iron composition range from low to high. In each composition, the iron atoms are placed in different metal cation site to build the different bulk structure.We named the structure with the cations in the M1 site(Figure. \ref{fig:Fo50Bulk} and Figure. \ref{fig:BulkS}).

The catalytic activity of olivines in biomass gasification was proposed to be due to the presence of iron atoms on the surface and tested in several experimental studies\cite{RN1921}. We construct a series iron contained olivine surface slabs based on two types (Figure. \ref{fig:BulkS}) most possible exposed surface terminations one type with $M2$ site exposed and another type with $M1$ site exposed based on our previous studies\cite{Geng2019a}.The $M2$ exposed type of surfaces are relative more flat than the $M1$ exposed ones, no matter the iron atoms are on the top or not. We only choose the low iron content (2 Fe atom/slab and 4 Fe atom/slab of metal atoms) in our slab models, so we can consider iron as a impurity so the relative stability of surfaces will not differ much from the forsterite slabs. This strategy fits the fact that natural olivine usually has relative low iron content. Therefore, the slab model we chose is good enough to represent the most likely stable surfaces. To minimize the effect of the periodic boundary conditions, a 30 \AA thick vacuum space is add on the top of the surface slab models. All the atoms in the slabs are relaxed prior to energy calculation. 

\subsection{Computation methods}

The DFT calculations make use of the Perdew-Burke-Ernzerhof (PBE) approximation to the exchange-correlation functional with a 700 eV kinetic energy cutoff in the plane wave basis set. The projector augmented wave method is used to describe the effect of the inner electrons. Monkhorst-Pack k-point mesh were used in the optimization of bulk crystal phase ($8 \times 8 \times 8$) and surface slabs ($8 \times 8 \times 1$).  The ground-state atomic geometries of the bulk and surface are obtained by minimizing the force on each atom to below 0.01 eV/\AA. The vienna Ab initio simulation package (VASP)\cite{RN1248,RN1250} is used to perform the DFT calculation. For minerals containing iron, the 3d orbitals of iron can make up by two sets of orbitals $t_{2g}$ and $e_g$. $Fe^{2+}$ can have high spin and low spin state, high-spin is the quintet ($S=2, 2S+1=5$) which the difference of electrons spin up and spin down is 4 per $Fe$ atom, while low-spin is the singlet ($S=0, 2S+1=1$) which the difference of electron spin up and spin down is 0 per $Fe$ atom. So spin polarization should be considered in the calculation due to the possible spin transition in many iron containing mantle minerals\cite{RN1833,RN1841}. All the calculation are run as spin-polarized and the ground state of the bulk system are tested with both DFT+U method with a effective U value (U=4.5 eV, J=0.9 eV) adopted from previous works\cite{RN1805,RN1806,RN1807} and strongly constrained appropriately normed (SCAN) meta-GGA\cite{RN1990}. As the SCAN functional sometime has some difficulties in geometry optimization, we used the optimized results with PBE as the initial structure of SCAN and PBE+U optimization calculation. The distribution of the Fe atom will not always be symmetry in the slab models. A dipole correction is applied in the slab calculation. 

\subsection{Thermodynamics}

In most studies, ground state total energies from DFT calculations were directly used for estimating the stability of different structures. But many experiments reported the ordering distribution of cations can affected by temperature. Iron bearing olivine will have order-disorder transition at certain temperature. Unfortunately the transition temperature also didn't reach an agreement in different experiments. By evaluating of the Gibbs free energy, the temperature and composition differences can be takeing into account in the analysis of structure stability. The Gibbs free energy $G$ can be calculated as 
\begin{equation}
	\begin{aligned}
		G(T)=E_{DFT}+F^{vib}(T)-TS^{conf}+pV
	\end{aligned}	
\end{equation}
$E_{DFT}$ is the ground state total energy from the DFT calculation, $F^{vib}(T)$ is the vibrational energy contribution, $S^{conf}$ is the configurational ("mixing") entropy. The volume per unit crystal cell is 299.12 \AA$^3$, from some experiment result the volume variation under 15GPa is small, which make the $pV$ term can be at the common error level of DFT calculation. On the another hand, the energy difference ($\Delta G$) of different structures is important in the comparison, the volume difference of the structures can be extreme small when the composition is the same, so we can neglected the $pV$ term. 
\begin{equation}
	\Delta G(T) = \Delta E+\Delta F^{vib}(T)-T\Delta S^{conf}
\end{equation}  
We considered the configurational entropy ($S^{conf}$) of the system explicitly as follow due to the different distribution of iron atoms in the lattice. 
\begin{equation}
	S^{conf}=-k_B\sum_j( m_j \sum _i X_{ij}ln X_{ij}) 
\end{equation}
where $m_j$ is the total number of atoms in the $j^{th}$ crystallographic and $X_{i,j}$ is the mole fraction of the $i^{th}$ atom in the $j^{th}$ site. 

The vibrational contribution term $F^{vib}(T)$ can be expressed using phonon density of states as follow
\begin{equation}
	\begin{aligned}
		F^{vib}(T)=&\frac{1}{2} \sum_{qj} h\omega _{qj} \\
		&+ k_BT\sum _{qj} ln[1-e^{(-h\omega _{qj})/k_BT}] 
	\end{aligned}
\end{equation}
where $q$ is the wave vector, $j$ is the band index, and $\omega_{qj}$ is the phonon frequency of the phonon mode labeled by set ${q,j}$. The phonon calculation are conducted using density functional perturbation theory (DFPT) as implemented in the VASP/Phonopy software \cite{RN1107}. 

Based on experience in a study of diffusion in solid system\cite{RN2051}, we used a model based on the energetics of swapping Mg and Fe atoms in inequivalent sites of the system which allows us to predicted the probability of Fe diffusion to the certain site. In this case, the metal sites can be occupied either by Mg atom or Fe atom. The probability of a site is occupied by Fe can be estimated by a Fermi-Dirac distribution model.
\begin{equation}
	\begin{aligned}
		n(\Delta E, \Delta\mu, T)=\frac{1}{e^{(\Delta E - \Delta\mu)/k_BT} + 1}
	\end{aligned}
\end{equation}
which $\Delta E$ is the swapping energy of the Mg and Fe atoms, $\Delta\mu$ is a relative chemical potential for replacing a Mg atom to Fe atom. Since the $\Delta\mu$ can be approximately estimated from the free energy difference of bulk system with one more Fe atom replacing the Mg atom to the reference system. But the Gibbs free energy from this approach  of the this approach is composition related, one atom change in a small system can lead large composition change. To overcome this, we calculated $\Delta\mu$ for a series of supercells have similar composition ratio as our surface slabs.    

\section{III. Results and Discussion}

\subsection{Structure optimization}

We have performed structural optimization for bulk systems of 3 different chemical compositions in order to obtain the most stable configuration as the initial crystal structures are adopted from the experiment result have a different chemical composition. With the Fe atoms added, the energy surfaces became rougher than the forsterite (pure magnesium structure) system, takes our some effort to reach the convergence in structure optimization. 

The optimized results of the primitive cell of olivine with different composition in orthorhombic structures are shown in Table \ref{tab:LatticeConstant}. There is no experimental data available with the exact same chemical composition for some structures. Only the Fo50 structure can be directly compared to the experimental result. However, by comparing the experimental results \cite{RN1981,RN1830,RN1982} of pure end members forsterite, fayalite and Fo50, the error in our results should be less than 5\%. 

As shown in Table. \ref{tab:PolyVolume}, no matter in the forsterite or fayalite structure, the M1 octahedra is slightly smaller than the M2 octahedra. So some previous studies claims the slight preference of the Fe atom to M1 site is because of the ion radius of $Fe^{2+}$ is smaller than $Mg^{2+}$\cite{RN1992}.However, the ion radius are not the same between two spin states. The $Fe^{2+}$ radius is smaller than $Mg^{2+}$ only when the $Fe^{2+}$ is in its low spin state\cite{RN1968}. Usually the low spin state is the dominate state under high pressure when the orbital splitting is smaller. Theoretic research\cite{RN1917} predicted the spin transition from high spin to low spin of the Fayalite will happen around 15GPa, and experimental observation\cite{RN1886} suggested this spin transition would happen under higher pressure, between 40 and 50 GPa. We evaluated the electron-spin multiplicity by fix the difference of electron spin up and spin down in the VASP calculation. Our calculation result for bulk in Table. \ref{tab: BulkSpin} shows high spin states always have a lower energy than low spin. Our calculation result fits the experiment fact that the $Fe^{2+}$ is in the high spin state where natural olivine exists. The ion radius of high spin $Fe^{2+}$ is bigger, so the octahedra need larger space. Correspondingly in the Table. \ref{tab:PolyVolume}, we found the octahedra will be expanded when iron in the site, regardless in M1 or M2 site.  In all the conditions, $SiO$ tetrahedra is slight larger than the tetrahedra in forsterite (pure magnesium) or fayalite (pure iron), but when the composition fixed the tetrahedra will keep about the same size.

The slab results in Table. \ref{tab:SlabSpin} show high spin states $Fe^{2+}$ also have the lower energies in surface slabs. The $Fe^{2+}$ should occupy larger space than $Mg^{2+}$. However, the surface geometry feature breaks some polyhedra in the surface slab. Grid based Bader analysis\cite{RN2010} (Table \ref{tab:Bader}) gave us the atomic volume of the cations purely based on the electronic charge density. We find the Fe atoms always have a larger volume than Mg no matter inside the bulk or exposed on the surface. The Fe atoms involvement did not change the charge distribution of the slab. Similar to our previous result with forsterite surface, the charge differences of the same atom is small in the slabs. But the oxygen atoms are divided into groups by the cation it associated with.   

\subsection{Ground state preference: Total-energy calculations}

To test the calculation level of our study, we calculated the total energies of Fo50 ($Mg:Fe=4:4$) olivines within PBE, PBE+U and SCAN three different functionals (Table. \ref{tab:Total-Energy-44}). We find the energy of the system increases with the swapping of the Fe atoms from M1 site to M2 site. In PBE calculation, the energy difference between all $Fe^{2+}$ in the M1 and all $Fe^{2+}$ in the M2 is $0.475 eV$ and the atom difference between is 4 atoms. Some previous studies used GGA+U to calculate the elasticity of fayalite and hydrous fayalite due to the 3d electrons of Fe atom\cite{RN1807}. In VASP the U value is setted by two constant U and J. We simply adopted the U value from some previous studies\cite{RN1805, RN1807,RN1895} to test energies of the structures in our study since different studies used the same U and J value. The results we get from PBE+U calculation did not show enough difference whether the Fe atoms are in M1 or M2 site. At same time we used the SCAN which recently developed meta-GGA functional shows good performance on calculating system with transition metal atoms. We get a similar energy difference ($0.314 eV$) as PBE calculation in the SCAN calculation.The energies in SCAN calculation show the same trend as the PBE result, all the Fe atom in the M1 site has the lowest energy among comparing structures. And both the energies from PBE and SCAN results increases with the more iron atom moved to M1 site from M2 site and it grows linearly with the atom move which indicated Fe atoms prefers the M1 site although the $Fe^{2+}$ cations in high spin state have the larger ion radius than $Mg^{2+}$. The adopted U value in our study is not good enough to find out the energy difference of the distribution variation of the olivine system. The U value should be adjusted when the composition changes. From the results above, we find the PBE functional is good enough for find out the energy preference of different configuration, and also the structures we investigate are not the same in chemical composition. So we used PBE for the following calculation.

A similar energy characteristic shows in both the high and low Fe concentrations conditions. The more Fe atom in the M1 site the lower energy the system have. When the Fe concentration drops to $Mg:Fe=6:2$ (Fo75), the highest difference of Fe in M1 site and M2 site is 2 Fe atoms per unitcell, and the the energy difference is $0.212 eV$. In Fo25, the structures have higher Fe concentration ($Mg:Fe=2:6$). The largest Fe atom difference between structures is also 2 atoms per unitcell which have a energy difference of $0.492 eV$. We find the energy difference per atom keeps about the same from Fo75 to Fo50 and increases from Fo50 to Fo25 with the Fe concentration raise. This indicate the more Fe in the system the higher energy preference to move the Fe atom from M2 to M1 sites. 

\subsection{Entropy effect}

Recently study\cite{RN1927} on Fe containing carbonate claimed the magnetic entropy caused by the distribution of Fe atoms with different spin state can be important in mineral stabilization in the mantle. Since the natural occurrence of olivine is mainly no more than 14GPa, far lower than the experiment observed spin transition pressure, we did not consider the systems under higher pressure. As described before, high spin states always have a lower energy in our calculation, which means all the Fe atoms are all in the same spin state, so the magnetic entropy could be neglected in our research. 

\begin{figure*}[h]
	\centering
	\includegraphics[width=0.9\linewidth]{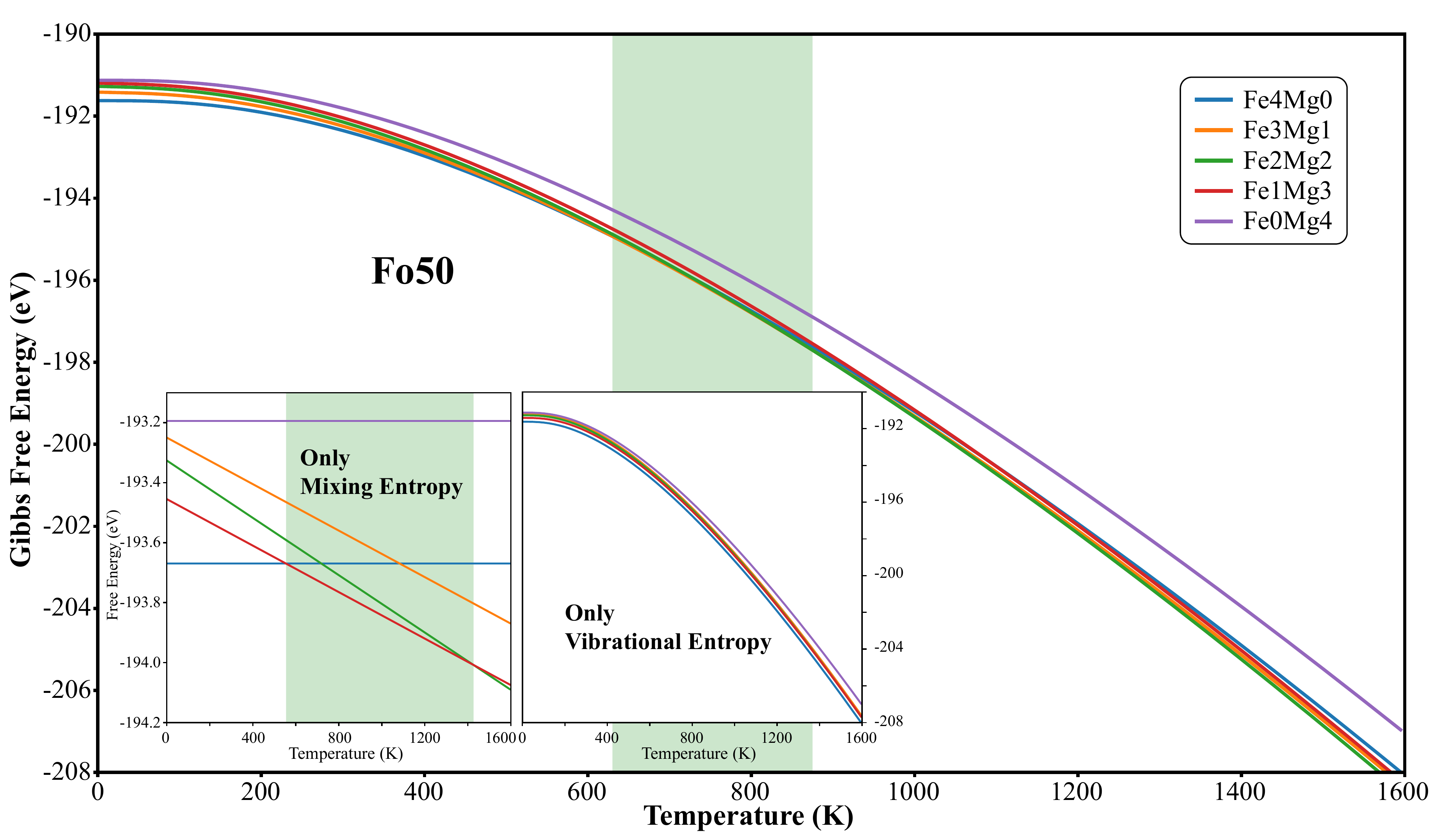}
	\caption{Free energy comparison with different entropy contributions of Fo50. The shade span on the background is the energy crossover temperature range. Temperature below the crossover range the M1 ordered structure (Fe4Mg0) is the most stable, and above this range the fully mixed structure (Fe2Mg2) is most stable.Two Ordered structures have 0 mixing entropy contribution, so the energies are flat in the only mixing entropy graph. Only consider vibrational entropy the M1 ordered structure (Fe4Mg0) will always be the most stable structure.}
	\label{fig:Mixing50}
\end{figure*}

Since two different kinds of cations are involved in the system, the structures can be category into two groups: ordered group and mixed group.  For example in Figure. \ref{fig:Fo50Bulk}, Fo50 $Fe0Mg4$ and $Fe4Mg0$ are the ordered structures which have the lowest and highest $k_D$ value respectively. The $k_D$ value of $Fe4Mg4$ in Fo50 which has the fully mixed distribution of $Mg$ and $Fe^{2+}$ cations is 1. These cation distributions cause the 0K total energy difference due to the preference of $Fe^{2+}$. The more $Fe^{2+}$ distributed in the M1 site, the lower energy the structure have. The ordered group have the lowest and highest energy and the mixed group lays in between the two ordered structures. However, these two type of cations distributed into different sites will cause the mixing entropy. The mixing entropy is much less than $-0.01 eV/K$ and the free energy contribution is just meV level when at the low temperature. Pressure effects on the cation distribution order-disorder of main minerals was reviewed by Hazen and Navrotsky\cite{RN1901}. $\Delta V_{dis}= V_{disordered}-V_{ordered}$ derived from Akamatsu et al.\cite{RN1983} experiment is only $\Delta V_{dis}=0.24 cm^3/mol=1.6 \times 10^{-3} \AA^3/unitcell$, so we can neglected the pressure effect of the mixing energy contribution in our pressure range. When temperature raise to 1500K, the maximum difference of free energy contribution of mixing entropy between orderer and disordered structure is about $0.79 eV$ which is about the similar order of magnitude of the ground total energy difference caused by the $Fe^{2+}$ distribution. Therefore, the mixing entropy should be considered in the free energy when evaluating the temperature effect of the cation distribution. Vibrational entropy is also believed to play a important role in free energy contribution at high temperature. As we can find in the Figure. \ref{fig:Mixing50}, the vibrational entropy only shrunken the energy gaps of the mixed group structures and the M1 ordered structure and no energy crossover is found under 1600K in our study. The mixing entropy only can cause some energy crossovers in the temperature range. 

\begin{figure*}[h]
	\centering
	\includegraphics[width=0.9\linewidth]{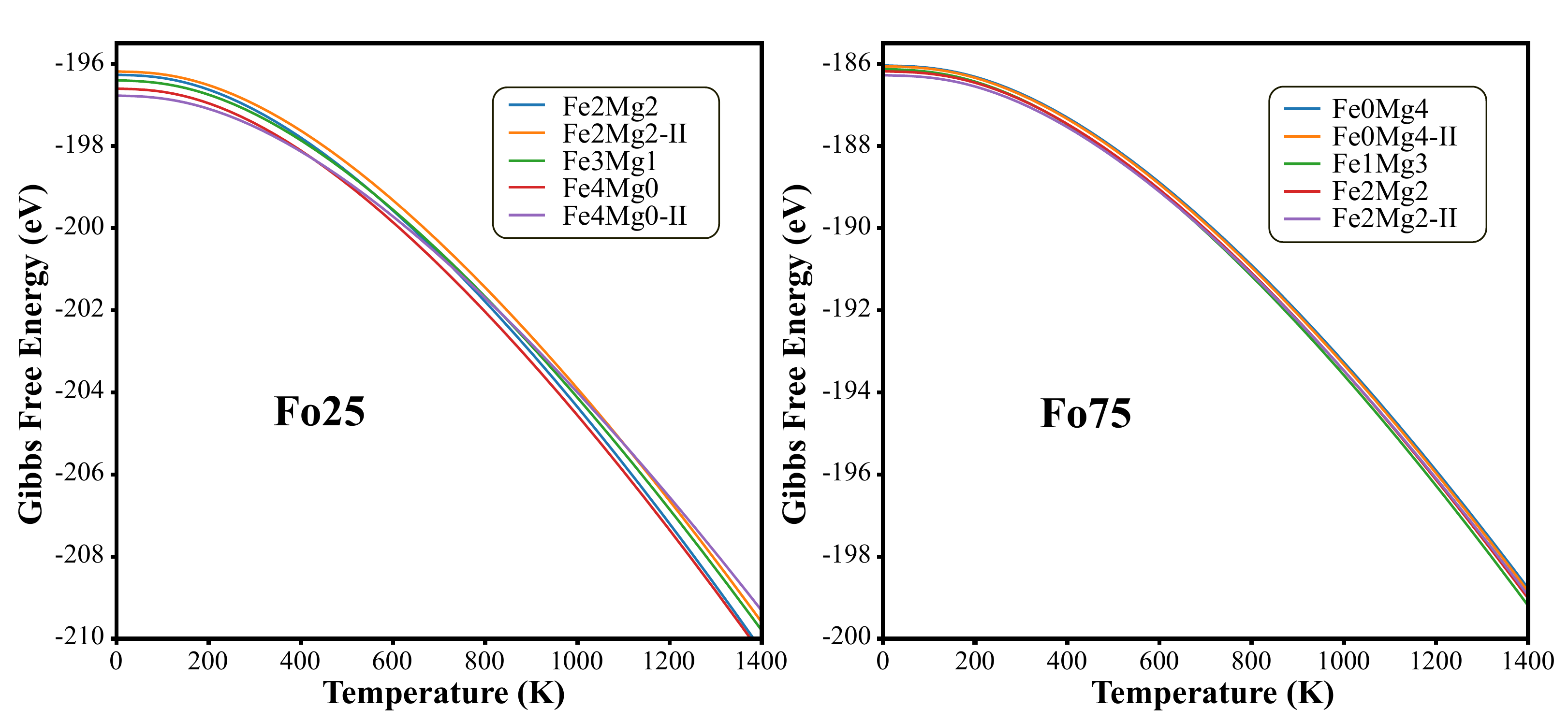}
	\caption{Gibbs free energy comparisons of Fo25 and Fo75 structures.The crossover temperature changes with the composition variation, lower Fe content have lower crossover temperature and the structure which have the $k_D=1$ reaches the most stable above the crossover temperature.}
	\label{fig:Gibbs25-75}
\end{figure*}

As we expected, with both mixing and vibrational entropy considered in the Gibbs free energy evaluation, we can find the fully ordered structure $Fe0Mg4$ is the most stable structure at low temperature and the most disordered structure $Fe2Mg2$ is the most stable structure at the high temperature. Except the ordered M2 structure $Fe4Mg0$, the energy difference is small among the rest structures. The large energy gap between M2 ordered structure and the rest also indicate it is highly difficult to distribute $Fe^{2+}$ into M2 site even at high temperature. In experiment, it is quite difficult to precisely know one site is taken over by $Fe^{2+}$ or $Mg$ \cite{RN1825}, so $k_D$ value became their important indicator to know the site preference. In Fo50 by Redfern et al.\cite{RN1830} find out the $k_D = 1$ at around $898 K $. We can find a similar energy crossover of the most disorder structure $Fe2Mg2$ which $k_D = 1$ in Fo50 at $850 K$ and became the most stable one among 5 possible structures. This crossover also means the site preference of $Fe^{2+}$ to M1 will disappear above this temperature, unlike many experiment research claimed the preference will increases continuously with rising temperature\cite{RN1906,RN1897}. Figure. \ref{fig:Gibbs25-75} shows this crossover temperature moves to lower temperature when the chemical composition is more $Mg$ rich and $Fe$ poor, and the $k_D$ value in different chemical system is also not the same. 

\subsection{Surface Exposure}

\begin{figure*}[h]
	\centering
	\includegraphics[width=0.9\linewidth]{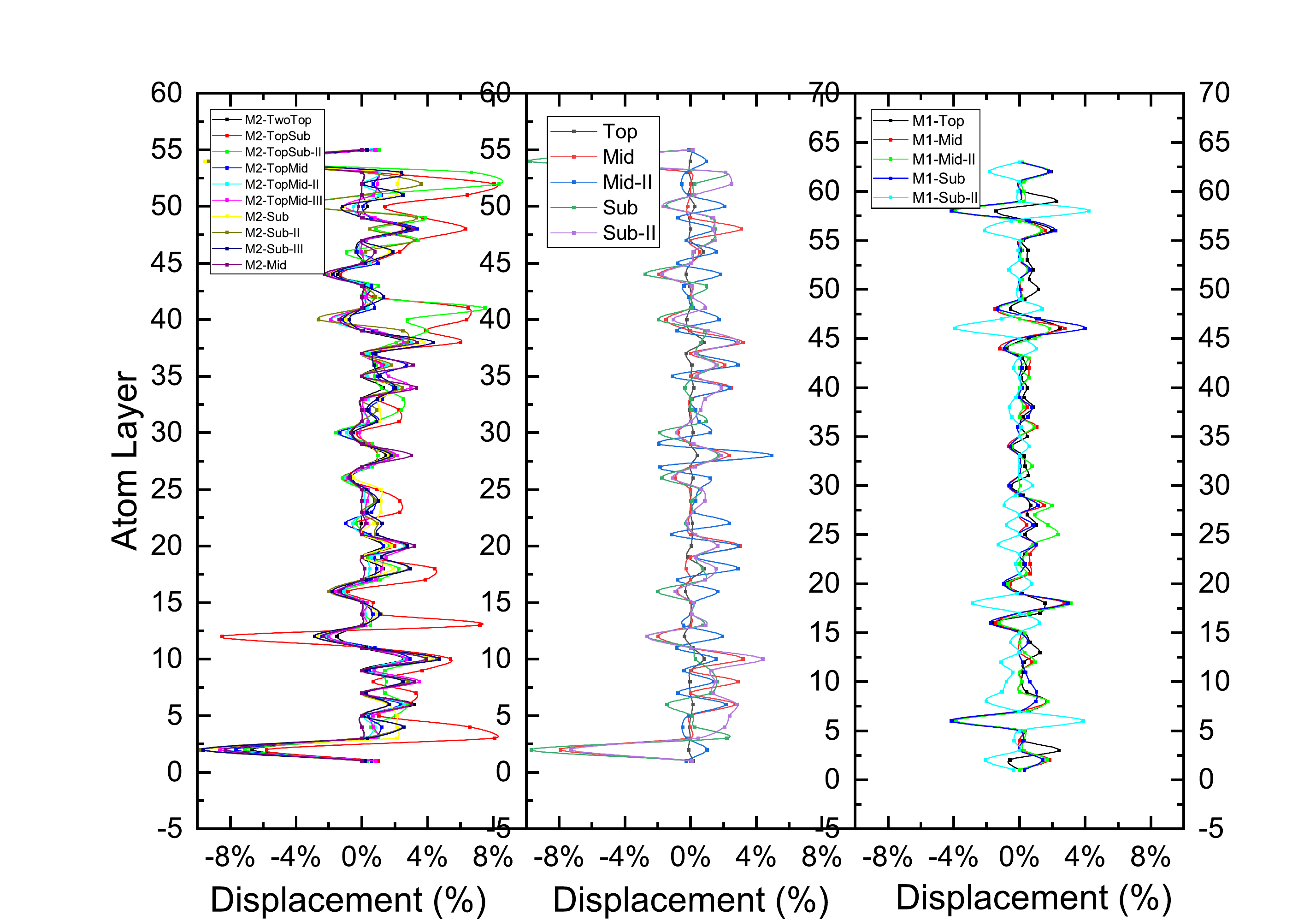}
	\caption{Atom displacements between origin cleaved slab and relaxed slabs grouped by the different surface termination and Fe concentration. The displacement $d = \frac{\Delta Z^{origin}_{i,j}-\Delta Z^{relaxed}_{i,j}}{\Delta Z_0}$, $\Delta Z^{origin}_{i,j}$ is the distance between atoms i and j in unrelaxed slab, $\Delta Z^{relaxed}_{i,j}$ is the distance between atoms i and j in relaxed slab, $\Delta Z_0$ is the length of thickness of the surface slab.}
	\label{fig:Displacement}
\end{figure*}

Although olivine was used as catalyst in many chemical reactions, very limited research were done on the surface cation distribution. In the surface slab, the surface exposure of a site can be quite critical to the cation distribution due to the unique geometric characteristic of the surface. The atoms exposed on the surface will gain extra space from the outside which can be seen from the Figure. \ref{fig:Displacement}. Atoms near the surface have larger displacements than the atom inside the slab during the relaxation. Similar to what we found in the bulk system result, the $Fe^{2+}$ in high spin state also have lower energy than in the low spin state on the surface site. Previous study on Fo75 olivine surface with B3LYP functional showed the quintet state is the lowest energy state\cite{RN1449} when the $Fe^{2+}$ occupied the surface site. But we did not find the $Fe^{2+}$ have larger splitting of $3d$ orbitals under saturated environments inside the slab their study reported\cite{RN1449}. In bulk system the $Fe^{2+}$ is always in saturated environment and the spin transition from high spin to low spin only happens under high pressure. In our result, the $Fe^{2+}$ in the slab showed similar behavior as it in the bulk system, quintet state always has the lowest energy no matter it is inside the slab or exposed on the surface. 

A preference to M1 site than M2 site of $Fe^{2+}$ can be observed if all $Fe^{2+}$ atom is inside the slab. The relative position of $Fe^{2+}$s also affect the energies. Structures with two $Fe^{2+}$ cations next to each other have lower energy than the structures two $Fe^{2+}$ separated by $Mg^{2+}$. However, the metal site exposed on the surface can lead a obvious preference of $Fe^{2+}$ regardless the site is M1 or M2 site. But the energy needed to swap $Fe^{2+}$ to the surface is not the same because the different cleavage energy to form different surface terminations. As we mentioned before, the octahedra of the metal site would be enlarged when $Fe^{2+}$ occupied the site. Obviously, the on surface exposed site can easily provide larger space than any other metal site inside the slab for the $Fe^{2+}$ because of the unsaturated geometric nature. The energy difference of surface exposure cation site and any interior cation site will highly possible become a driving force for $Fe^{2+}$ diffusion from any cation site to the surface. 

\begin{figure*}[h]
	\centering
	\includegraphics[width=0.9\linewidth]{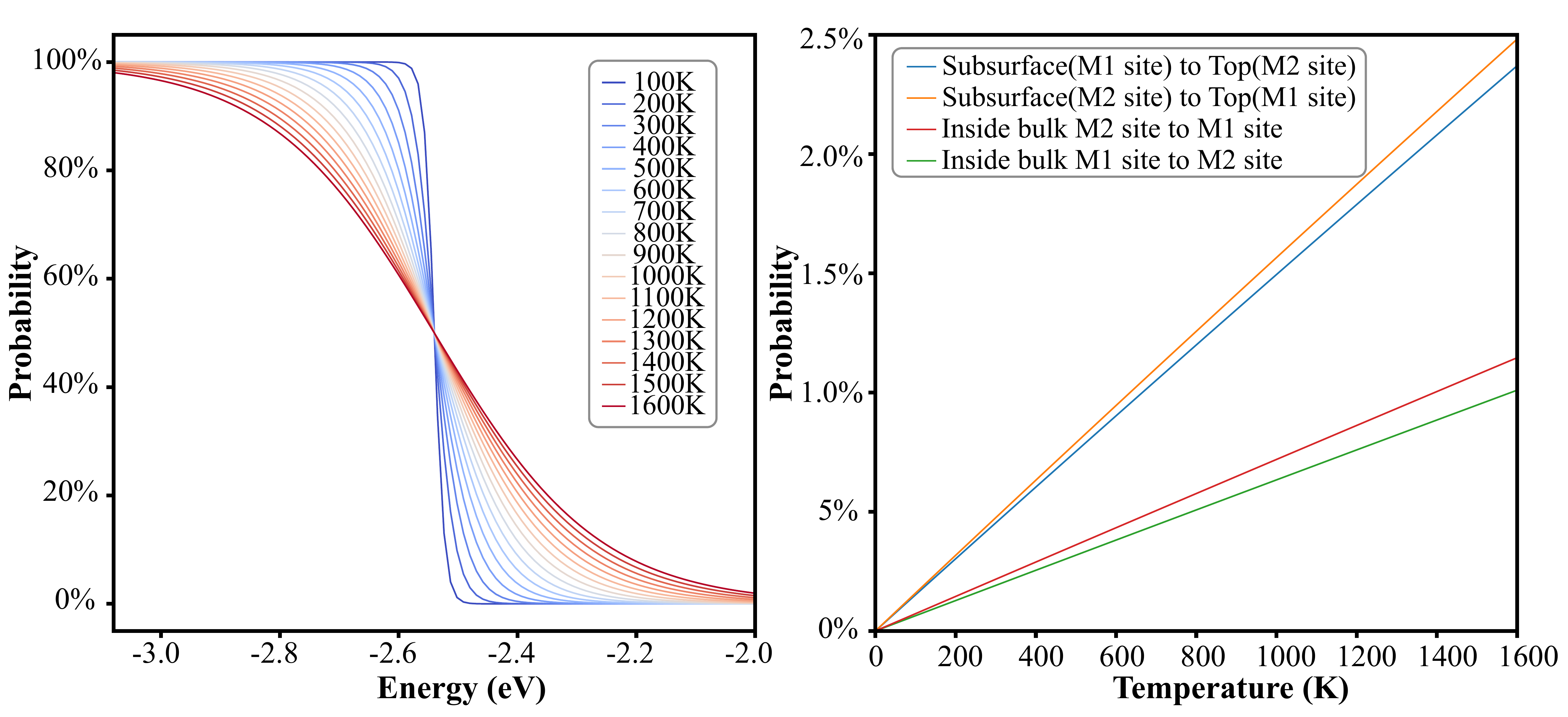}
	\caption{Occupation probability estimation from a Fermi-Dirac distribution based on the energetics. We got the relative chemical potential of changing a Mg atom to a Fe atom from supercells results. From the estimation we can find the higher probability for Fe atoms moving form M2 site to M1 site in the bulk, and even higher probability if the Fe atoms is moving form subsurface layer to the exposed metal site regardless the type of the site. This would lead the surface have higher Fe concentration than the bulk system after the diffusion of the Fe atoms. }
	\label{fig:Probability}
\end{figure*}

With the relative chemical potential get from the supercell results, we estimated the occupation probability of Fe atoms move to particular site after diffusion based on the energetics in Figure \ref{fig:Probability}. We compared the probability of moving $Fe^{2+}$ to M2 site inside the bulk and if the M2 site is on the surface. It is quite clear that $Fe^{2+}$ shows higher probability move to M2 site if the site is exposed on the surface. This also indicated the surface of the olivine can have a higher Fe concentration than the bulk due to the diffusion of Fe atom and the catalytic ability iron contained olivine shown in many reactions is highly based on the $Fe^{2+}$ on the surface.A lot of natural olivine have a kind of zoning structure which have higher Fe concentration on the edge than the center of the crystal. The chemical zoning of the natural olivine was think formed by the sequential magma activity when the olivine growing, but the diffusion of metal ions after the olivine crystallized became more focused and became an emerging tool to understand the volcano's eruptive past\cite{RN2103}. Since the $Fe^{2+}$'s trend to diffuse to the surface which behave the same pattern as the experiment observed in the olivine chemical zoning. Grain boundary effect which is ignored in the past models\cite{RN2098} should be definitely considered in the future models of diffusion chronometry. 

\section{IV. Conclusion}

We used DFT calculation combined thermodynamics to analyze bulk crystal and surface slab of olivine with different Fe concentration from iron rich (Fo25) to iron poor (Fo75). The relative stability of olivine crystals and slabs with different $Fe^{2+}$ cation distribution is estimated based on the Gibbs free energy. Meanwhile, this relative stability indicated the preference of Fe atom between different metal sites in the olivine crystal structure. We found the high spin state is the most stable spin state of $Fe^{2+}$ under ambient pressure, which lead a larger ion radius of $Fe^{2+}$ than $Mg^{2+}$ in the crystal. This means a larger metal-oxygen octahedra is needed in olivine to put the $Fe^{2+}$ in. Although the M1 site octahedra is smaller than M2 site in both Mg and Fe endmembers of olivine, the $Fe^{2+}$ is still energetically  prefer M1 site, from both our PBE and SCAN calculations, when Fe is added to the system by enlarge the metal-oxygen octahedra. This is not consistent with previous study that $Fe^{2+}$ prefer M1 site because of its smaller size than $Mg^{2+}$. The energy contribution of entropies accumulating with the temperature increases, and the vibrational entropy shrink the energy difference of different structures while the mixing entropy fills the energy gap of order and disorder distribution of $Fe^{2+}$ in the olivine crystal under high temperature at the same time. At the order-disorder energy crossover temperature, the preference to M1 site of $Fe^{2+}$ will disappear due to these entropy contributions.    

Similar to the bulk system, high spin state also is the stable state of $Fe^{2+}$ in surface slabs. This lead $Fe^{2+}$ inside the surface slab distribute with the same principle as in the bulk. Once a metal site expose on the surface will easily provide larger space for the cation because of the geometric character. This feature make surface exposed metal site became more preferable to $Fe^{2+}$ than any metal site inside the slab no matter M1 or M2 site. This indicate the impurity $Fe^{2+}$ in olivine can be highly possible on the surface of olivine by diffusion from bulk to the surface driving by the energy preference. Since $Fe^{2+}$ have a $3d$ electron orbital, this implication shows Fe atom in the olivine can play a critical role as a catalyst in many geochemical reactions which many are shown in the olivine related reactions.

%%%%%%%%%%%%%%%%%%%%%%%%%%%%%%%%%%%%%%%%%%%%%%%%%%%%%%%%%%%%%%%%%%%%%
%% The "Acknowledgement" section can be given in all manuscript
%% classes.  This should be given within the "acknowledgement"
%% environment, which will make the correct section or running title.
%%%%%%%%%%%%%%%%%%%%%%%%%%%%%%%%%%%%%%%%%%%%%%%%%%%%%%%%%%%%%%%%%%%%%
\begin{acknowledgement}

This work was supported by ???.
The computations were performed on resources provided by the Icelandic High Performance Computing Centre at the University of Iceland.

\end{acknowledgement}

%%%%%%%%%%%%%%%%%%%%%%%%%%%%%%%%%%%%%%%%%%%%%%%%%%%%%%%%%%%%%%%%%%%%%
%% The same is true for Supporting Information, which should use the
%% suppinfo environment.
%%%%%%%%%%%%%%%%%%%%%%%%%%%%%%%%%%%%%%%%%%%%%%%%%%%%%%%%%%%%%%%%%%%%%
%\begin{suppinfo}

%\end{suppinfo}

%%%%%%%%%%%%%%%%%%%%%%%%%%%%%%%%%%%%%%%%%%%%%%%%%%%%%%%%%%%%%%%%%%%%%
%% The appropriate \bibliography command should be placed here.
%% Notice that the class file automatically sets \bibliographystyle
%% and also names the section correctly.
%%%%%%%%%%%%%%%%%%%%%%%%%%%%%%%%%%%%%%%%%%%%%%%%%%%%%%%%%%%%%%%%%%%%%
\bibliography{Manuscript}

\clearpage

\begin{table*}[h]
	\caption{Experimental and Calculated Values for Cell Parameters of Bulk Forsterite (in \AA)}
	\label{tab:LatticeConstant}	
	\begin{tabular}{c|c|c|c|c|c|c}
		\hline\hline
	    \multirow{2}{*}{\begin{tabular}[c]{@{}c@{}}Lattice\\ Constant\end{tabular}} & \multicolumn{3}{c|}{Expriment} & \multicolumn{3}{c}{This work} \\ \cline{2-7} 
		  & Forsterite \cite{RN1981} & Fo50 \cite{RN1830}  & Fayalite\cite{RN1982} & Fo75  & Fo50  & Fo25  \\ \hline
		a & 4.76   & 4.81  & 4.82   & 4.83  & 4.88  & 4.88  \\ 
		b & 10.20  & 10.38 & 10.49  & 10.50 & 10.51 & 10.51 \\ 
		c & 5.98   & 6.06  & 6.10   & 6.14  & 6.16  & 6.16  \\ \hline\hline
		\end{tabular}
\end{table*}

\begin{table*}[h]
	\caption{Bader analysis of the surface slabs. The bader volume is in {\AA}$^3$ and the bader charge of different surface slabs. }
	\label{tab:Bader}
	\resizebox{\textwidth}{20mm}{
	\begin{tabular}{c|c|cccccccc|c|c|cccccc}
	\hline\hline
	\multicolumn{2}{c|}{\multirow{4}{*}{Surface Slab}} &
	  \multicolumn{8}{c|}{Bader Volume (Charge)} &
	  \multicolumn{2}{c|}{\multirow{4}{*}{Surface Slab}} &
	  \multicolumn{6}{c}{Bader Volume (Charge)} \\ \cline{3-10} \cline{13-18} 
	\multicolumn{2}{c|}{} &
	  \multicolumn{4}{c|}{Top} &
	  \multicolumn{4}{c|}{Mid} &
	  \multicolumn{2}{c|}{} &
	  \multicolumn{2}{c|}{Top} &
	  \multicolumn{4}{c}{Mid} \\ \cline{3-10} \cline{13-18} 
	\multicolumn{2}{c|}{} &
	  \multicolumn{2}{c|}{Fe} &
	  \multicolumn{2}{c|}{Mg} &
	  \multicolumn{2}{c|}{Fe} &
	  \multicolumn{2}{c|}{Mg} &
	  \multicolumn{2}{c|}{} &
	  \multicolumn{2}{c|}{M2} &
	  \multicolumn{2}{c|}{Fe} &
	  \multicolumn{2}{c}{Mg} \\ \cline{3-10} \cline{13-18} 
	\multicolumn{2}{c|}{} &
	  \multicolumn{1}{c|}{M1} &
	  \multicolumn{1}{c|}{M2} &
	  \multicolumn{1}{c|}{M1} &
	  \multicolumn{1}{c|}{M2} &
	  \multicolumn{1}{c|}{M1} &
	  \multicolumn{1}{c|}{M2} &
	  \multicolumn{1}{c|}{M1} &
	  M2 &
	  \multicolumn{2}{c|}{} &
	  \multicolumn{1}{c|}{Fe} &
	  \multicolumn{1}{c|}{Mg} &
	  \multicolumn{1}{c|}{M1} &
	  \multicolumn{1}{c|}{M2} &
	  \multicolumn{1}{c|}{M1} &
	  \multicolumn{1}{c}{M2} \\ \hline
	\multirow{5}{*}{\begin{tabular}[c]{@{}c@{}}M2 termination \\ ( 2 Fe Atoms )\end{tabular}} &
	  Top &
	   &
	  159.6 (1.22) &
	   &
	  106.8 (1.67) &
	   &
	   &
	  4.9 (1.68) &
	  5.3 (1.70) &
	  \multirow{10}{*}{\begin{tabular}[c]{@{}c@{}}M2 termination \\ ( 4 Fe Atoms )\end{tabular}} &
	  TwoTop &
	  162.3 (1.23) &
	   &
	   &
	   &
	  4.9 (1.68) &
	  5.3 (1.70) \\ \cline{2-2} \cline{12-12}
	 &
	  Sub &
	   &
	   &
	   &
	  92.8 (1.67) &
	  10.8 (1.32) &
	   &
	  4.9 (1.68) &
	  5.2 (1.70) &
	   &
	  Top-Sub &
	  90.1 (1.14) &
	  110.3 (1.67) &
	   &
	  11.1 (1.31) &
	  4.9 (1.68) &
	  5.3 (1.70) \\ \cline{2-2} \cline{12-12}
	 &
	  Sub-II &
	   &
	   &
	   &
	  93.0 (1.66) &
	  10.8 (1.32) &
	   &
	  4.9 (1.68) &
	  5.2 (1.70) &
	   &
	  Top-Sub-II &
	  125.7 (1.17) &
	  104.6 (1.67) &
	   &
	  10.9 (1.33) &
	  4.9 (1.68) &
	  5.3 (1.70) \\ \cline{2-2} \cline{12-12}
	 &
	  Mid &
	   &
	   &
	   &
	  94.8 (1.66) &
	   &
	  10.9 (1.38) &
	  4.9 (1.68) &
	  5.4 (1.70) &
	   &
	  Top-Mid &
	  157.2 (1.22) &
	  108.4 (1.67) &
	  10.9 (1.39) &
	   &
	  4.9 (1.68) &
	  5.2 (1.70) \\ \cline{2-2} \cline{12-12}
	 &
	  Mid-II &
	   &
	   &
	   &
	  93.4 (1.64) &
	   &
	  11.0 (1.38) &
	  4.9 (1.68) &
	  5.2 (1.70) &
	   &
	  Top-Mid-II &
	  157.7 (1.21) &
	  106.3 (1.67) &
	  10.8 (1.38) &
	   &
	  4.9 (1.68) &
	  5.3 (1.70) \\ \cline{1-2} \cline{12-12}
	\multirow{5}{*}{\begin{tabular}[c]{@{}c@{}}M1 termination\\ ( 2 Fe Atoms )\end{tabular}} &
	  Top &
	  69.0 (1.15) &
	   &
	  9.1 (1.65) &
	   &
	   &
	   &
	  4.9 (1.68) &
	  5.3 (1.70) &
	   &
	  Top-Mid-III &
	  158.1 (1.22) &
	  102.8 (1.67) &
	  10.9 (1.39) &
	   &
	  4.9 (1.68) &
	  5.2 (1.70) \\ \cline{2-2} \cline{12-12}
	 &
	  Sub &
	   &
	   &
	  8.7 (1.65) &
	   &
	   &
	  11.2 (1.35) &
	  4.9 (1.68) &
	  5.4 (1.70) &
	   &
	  Sub &
	   &
	  88.4 (1.67) &
	  10.7 (1.32) &
	   &
	  4.9 (1.68) &
	  5.2 (1.70) \\ \cline{2-2} \cline{12-12}
	 &
	  Sub-II &
	   &
	   &
	  8.8 (1.65) &
	   &
	   &
	  11.2 (1.35) &
	  4.9 (1.68) &
	  5.4 (1.70) &
	   &
	  Sub-II &
	   &
	  89.2 (1.67) &
	  10.8 (1.32) &
	   &
	  4.9 (1.68) &
	  5.3 (1.70) \\ \cline{2-2} \cline{12-12}
	 &
	  Mid &
	   &
	   &
	  9.3 (1.65) &
	   &
	  10.6 (1.33) &
	   &
	  5.0 (1.68) &
	  5.4 (1.70) &
	   &
	  Sub-III &
	   &
	  87.9 (1.67) &
	  10.8 (1.32) &
	   &
	  4.9 (1.68) &
	  5.3 (1.70) \\ \cline{2-2} \cline{12-12} 
	 &
	  Mid-II &
	  \multicolumn{1}{c}{} &
	  \multicolumn{1}{c}{} &
	  \multicolumn{1}{c}{9.1 (1.65)} &
	  \multicolumn{1}{c}{} &
	  \multicolumn{1}{c}{10.3 (1.31)} &
	  \multicolumn{1}{c}{} &
	  \multicolumn{1}{c}{5.0 (1.68)} &
	  5.4 (1.70) &
	   &
	  Mid &
	  \multicolumn{1}{c}{} &
	  \multicolumn{1}{c}{91.6 (1.67)} &
	  \multicolumn{1}{c}{} &
	  \multicolumn{1}{c}{11.0 (1.38)} &
	  \multicolumn{1}{c}{4.9 (1.68)} &
	  \multicolumn{1}{c}{} \\ \hline\hline
	\end{tabular}}
\end{table*}

\begin{table*}[h]
	\caption{Total energy of $FeMgSiO_4$ (Fe:Mg=4:4) in 0K with different functionals ($\Delta E= E_{Fe0Mg4}-E$)}
	\label{tab:Total-Energy-44}
	\resizebox{\textwidth}{10mm}{
	\begin{tabular}{c|c|c|c|c|c|c|c|c|c|c}
		\hline\hline
		& \multicolumn{2}{c|}{Fe0Mg4} & \multicolumn{2}{c|}{Fe1Mg3} & \multicolumn{2}{c|}{Fe2Mg2} & \multicolumn{2}{c|}{Fe3Mg1} & \multicolumn{2}{c}{Fe4Mg0}  \\
		\hline
		&Total Energy(E)& $\Delta E$&Total Energy(E)&$\Delta E$&Total Energy(E)&$\Delta E$&Total Energy(E)&$\Delta E$&Total Energy(E)&$\Delta E$  \\
		\hline
		PBE  &-193.195&0.00&-193.250&0.055&-193.326&0.131&-193.455&0.260&-193.669&0.475 \\
		SCAN &-300.161&0.00&-300.402&0.241&-300.420&0.259&-300.417&0.256&-300.475&0.314 \\
		PBE-U&-188.579&0.00&-189.960&1.381&-189.963&1.384&-189.944&1.365&-189.955&1.376 \\
		\hline\hline
	\end{tabular}}
\end{table*}

\begin{table*}[h]
	\caption{Polyhedron volume in Fo50, Forsterite and Fayalite ({\AA}$^3$). The Fe atoms form larger octahedra than Mg atom no matter in M1 or M2 site. SiO tetrahedra keep the similar size.}
	\label{tab:PolyVolume}
	\begin{tabular}{c|c|c|c|c|c|c|c|c}
		\hline\hline
		\multicolumn{2}{c|}{}                                                 & Fe0Mg4  & Fe1Mg3  & Fe2Mg2  & Fe3Mg1  & Fe4Mg0  & Forsterite & Fayalite \\ \hline
	    \multirow{2}{*}{\begin{tabular}[c]{@{}c@{}}M1\\Octahedra\end{tabular}}& Fe &         & 13.45 & 13.50 & 13.47 & 13.58 &            & 12.66  \\ \cline{2-2} 
																			  & Mg & 12.63 & 13.34 & 12.70  & 12.79 &         & 11.89    &          \\\cline{1-2}
		\multirow{2}{*}{\begin{tabular}[c]{@{}c@{}}M2\\Octahedra\end{tabular}}& Fe & 14.03 & 13.99 & 13.94 & 13.79 &         &            & 13.18  \\ \cline{2-2} 
																			  & Mg &         & 12.64  & 13.26 & 13.08 & 13.11 & 12.48    &          \\\cline{1-2}
		\multicolumn{2}{c|}{SiO tetrahedra}                                   & 2.32  & 2.32  & 2.32  & 2.32  & 2.34  & 2.28     & 2.20   \\ \hline\hline
		\end{tabular}
\end{table*}

\begin{table*}[h]
	\caption{Relative Energies of bulk structures in high spin and low spin (eV), Reference value is the high spin total energy of least Fe in the M1 site structure in the table for each chemical composition. }
	\label{tab: BulkSpin}
	\resizebox{\textwidth}{20mm}{
	\begin{tabular}{c|c|c|c|c|c|c|c|c|c|c}
	\hline\hline
	\multirow{3}{*}{\begin{tabular}[c]{@{}c@{}}Fe:Mg = 2:6\\Fo75\end{tabular}} &
	  \multicolumn{2}{c|}{Fe0Mg4} &
	  \multicolumn{2}{c|}{Fe0Mg4-II} &
	  \multicolumn{2}{c|}{Fe1Mg3} &
	  \multicolumn{2}{c|}{Fe2Mg2} &
	  \multicolumn{2}{c}{Fe2Mg2-II} \\ \cline{2-11} 
	 &
	  High Spin &
	  Low Spin &
	  High Spin &
	  Low Spin &
	  High Spin &
	  Low Spin &
	  High Spin &
	  Low Spin &
	  High Spin &
	  Low Spin \\ \cline{2-11} 
	 &
	   0.000  &
	   1.949  &
	   -0.019 &
	   2.069  &
	   -0.065 &
	   2.439  &
	   -0.128 &
	   2.909  &
	   -0.211 &
	  2.751   \\ \hline
	\multirow{3}{*}{\begin{tabular}[c]{@{}c@{}}Fe:Mg = 4:4\\ Fo50\end{tabular}} &
	  \multicolumn{2}{c|}{Fe0Mg4} &
	  \multicolumn{2}{c|}{Fe1Mg3} &
	  \multicolumn{2}{c|}{Fe2Mg2} &
	  \multicolumn{2}{c|}{Fe3Mg1} &
	  \multicolumn{2}{c}{Fe4Mg0} \\ \cline{2-11} 
	 &
	  High Spin &
	  Low Spin &
	  High Spin &
	  Low Spin &
	  High Spin &
	  Low Spin &
	  High Spin &
	  Low Spin &
	  High Spin &
	  Low Spin \\ \cline{2-11} 
	 &
	  0.000 &
	  5.904 &
	  -0.055 &
	  5.529 &
	  -0.131 &
	  2.913 &
	  -0.260 &
	  4.805 &
	  -0.474 &
	  4.407 \\ \hline
	\multirow{3}{*}{\begin{tabular}[c]{@{}c@{}}Fe:Mg = 6:2\\ Fo25\end{tabular}} &
	  \multicolumn{2}{c|}{Fe2Mg2} &
	  \multicolumn{2}{c|}{Fe2Mg2-II} &
	  \multicolumn{2}{c|}{Fe3Mg1} &
	  \multicolumn{2}{c|}{Fe4Mg0} &
	  \multicolumn{2}{c}{Fe4Mg0-II} \\ \cline{2-11} 
	 &
	  High Spin &
	  Low Spin &
	  High Spin &
	  Low Spin &
	  High Spin &
	  Low Spin &
	  High Spin &
	  Low Spin &
	  High Spin &
	  Low Spin \\ \cline{2-11} 
	 &
	  0.000 &
	  8.048 &
	  0.114 &
	  8.288 &
	  -0.106 &
	  7.610 &
	  -0.291 &
	   7.337&
	  -0.378 &
	  7.234\\ \hline\hline
	\end{tabular}}
\end{table*}

\begin{table*}[h]
	\caption{Relative Total Energies of surface slabs in high spin and low spin state (eV), Reference value is the high spin total energy of first structure in the table for each chemical composition.}
	\label{tab:SlabSpin}
	\resizebox{\textwidth}{25mm}{
	\begin{tabular}{c|c|c|c|c|c|c|c|c|c|c|c}
		\hline\hline
		\multirow{9}{*}{\begin{tabular}[c]{@{}l@{}}M2\\ termination\\ Slabs\end{tabular}} &
		  \multirow{3}{*}{\begin{tabular}[c]{@{}c@{}}Fe atoms=\\ 2\end{tabular}} &
		  \multicolumn{2}{c|}{Mid} &
		  \multicolumn{2}{c|}{Mid-II} &
		  \multicolumn{2}{c|}{Sub-II} &
		  \multicolumn{2}{c|}{Sub} &
		  \multicolumn{2}{c}{Top} \\ \cline{3-12} 
		 &                       & High Spin & Low Spin & High Spin & Low Spin & High Spin & Low Spin & High Spin & Low Spin & High Spin & Low Spin \\ \cline{3-12} 
		 &                       & 0.000  & 2.972 & -0.026  & 2.866 & -0.143  & 1.887 & -0.143  & 1.880 & -1.749  & 1.562 \\ \cline{2-12} 
		 &
		  \multirow{6}{*}{\begin{tabular}[c]{@{}c@{}}Fe atoms=\\ 4\end{tabular}} &
		  \multicolumn{2}{c|}{TwoTop} &
		  \multicolumn{2}{c|}{Top-Sub} &
		  \multicolumn{2}{c|}{Top-Sub-II} &
		  \multicolumn{2}{c|}{Top-Mid} &
		  \multicolumn{2}{c}{Top-Mid-II} \\ \cline{3-12} 
		 &                       & High Spin & Low Spin & High Spin & Low Spin & High Spin & Low Spin & High Spin & Low Spin & High Spin & Low Spin \\ \cline{3-12} 
		 &                       & 0.000  & -0.026 & 1.135  & 6.503 & 1.308  & 1.146 & 1.616  & 4.760 & 1.629  & 1.623 \\ \cline{3-12} 
		 &
		   &
		  \multicolumn{2}{c|}{Sub} &
		  \multicolumn{2}{c|}{Sub-II} &
		  \multicolumn{2}{c|}{Sub-III} &
		  \multicolumn{2}{c|}{Top-Mid-III} &
		  \multicolumn{2}{c}{Mid} \\ \cline{3-12} 
		 &                       & High Spin & Low Spin & High Spin & Low Spin & High Spin & Low Spin & High Spin & Low Spin & High Spin & Low Spin \\ \cline{3-12} 
		 &                       & 3.105  & 5.124 & 3.042  & 7.225 & 3.041  & 7.173 & 1.597  & 7.738 & 3.313  & 8.846 \\ \hline
		\multirow{3}{*}{\begin{tabular}[c]{@{}l@{}}M1\\ termination\\ Slabs\end{tabular}} &
		  \multicolumn{1}{c|}{\multirow{3}{*}{\begin{tabular}[c]{@{}c@{}}Fe atoms=\\ 2\end{tabular}}} &
		  \multicolumn{2}{c|}{Mid} &
		  \multicolumn{2}{c|}{Mid-II} &
		  \multicolumn{2}{c|}{Sub-II} &
		  \multicolumn{2}{c|}{Sub} &
		  \multicolumn{2}{c}{Top} \\ \cline{3-12} 
		 & \multicolumn{1}{c|}{} & High Spin & Low Spin & High Spin & Low Spin & High Spin & Low Spin & High Spin & Low Spin & High Spin & Low Spin \\ \cline{3-12} 
		 & \multicolumn{1}{c|}{} & 0.000  & 1.994 & -0.195  & 2.026 & -0.088  & -0.088 & -0.088  & 3.009 & -1.789  & 1.454 \\ \hline\hline
		\end{tabular}}
\end{table*}

\begin{figure*}[h]
	\centering
	\includegraphics[width=0.9\linewidth]{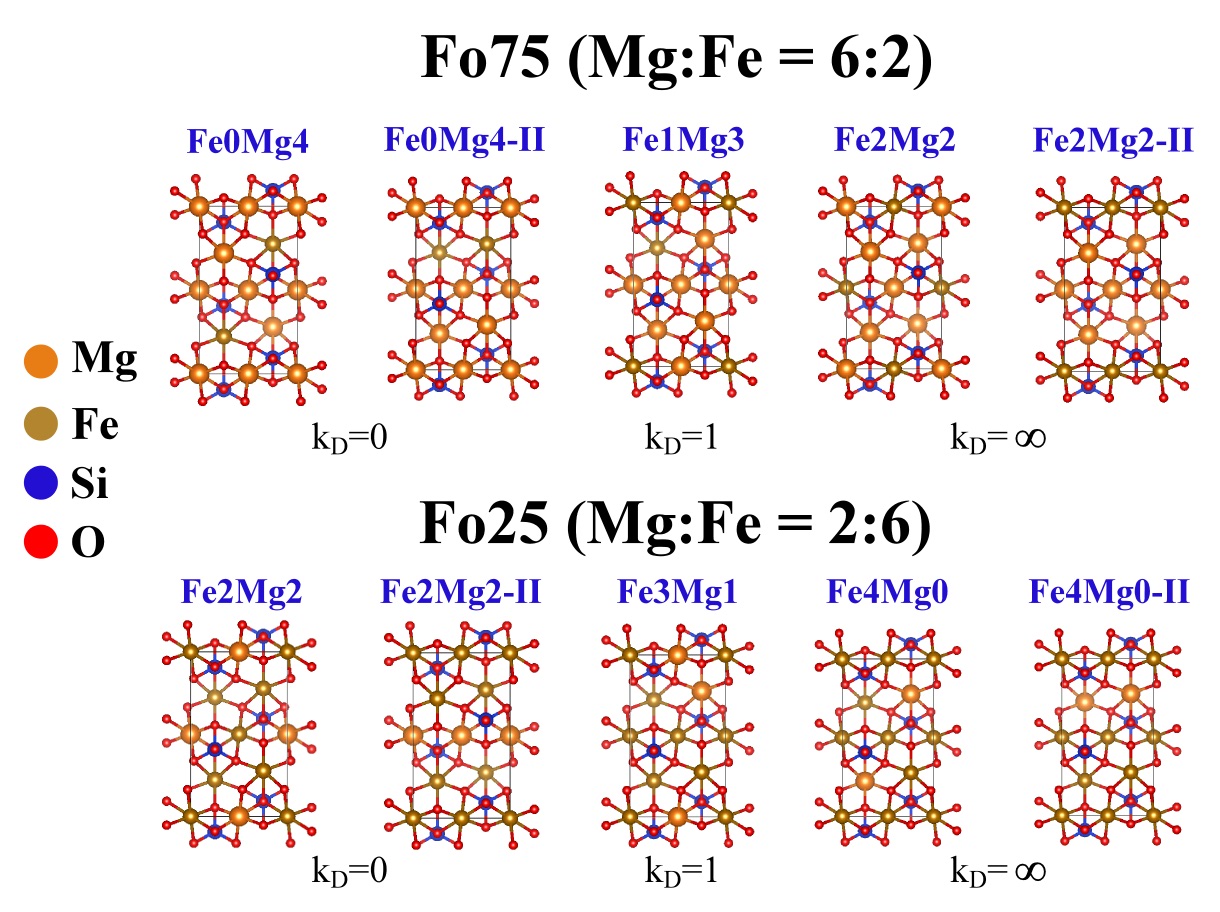}
	\caption{Structures of the $(Mg_xFe_{2-x})SiO_4$ with different Fe distribution in iron rich(Fo25) and iron poor(Fo75).The structure is named by the metal atom number in M1 sites. $k_D$ is the distribution coefficient. The $k_D$ value shift with the structure chemical composition. }
	\label{fig:BulkS}
\end{figure*}

\begin{figure*}[h]
	\centering
	\includegraphics[width=0.9\linewidth]{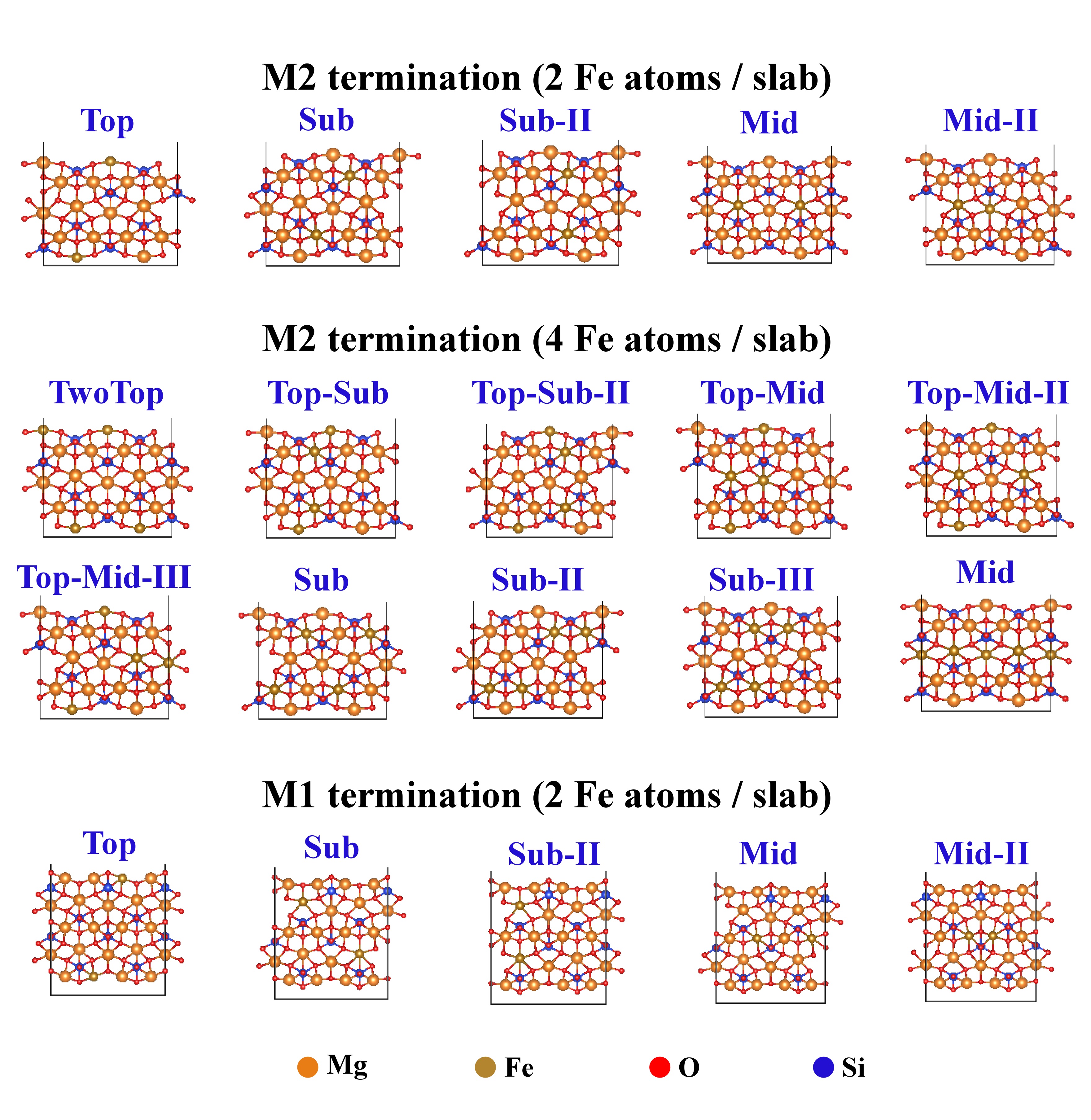}
	\caption{Side Views of structures of the slabs with different Fe distribution and different terminations. M2 termination and M1 termination mean the slab have M2 or M1 site exposed on the surface respectively. The naming is based on the Fe atom location of the slab. Top means on the surface, Sub means under the surface, and Mid means inside the interior of the slab.}
	\label{fig:BulkS}
\end{figure*}

\begin{figure*}[h]
	\centering
	\includegraphics[width=0.9\linewidth]{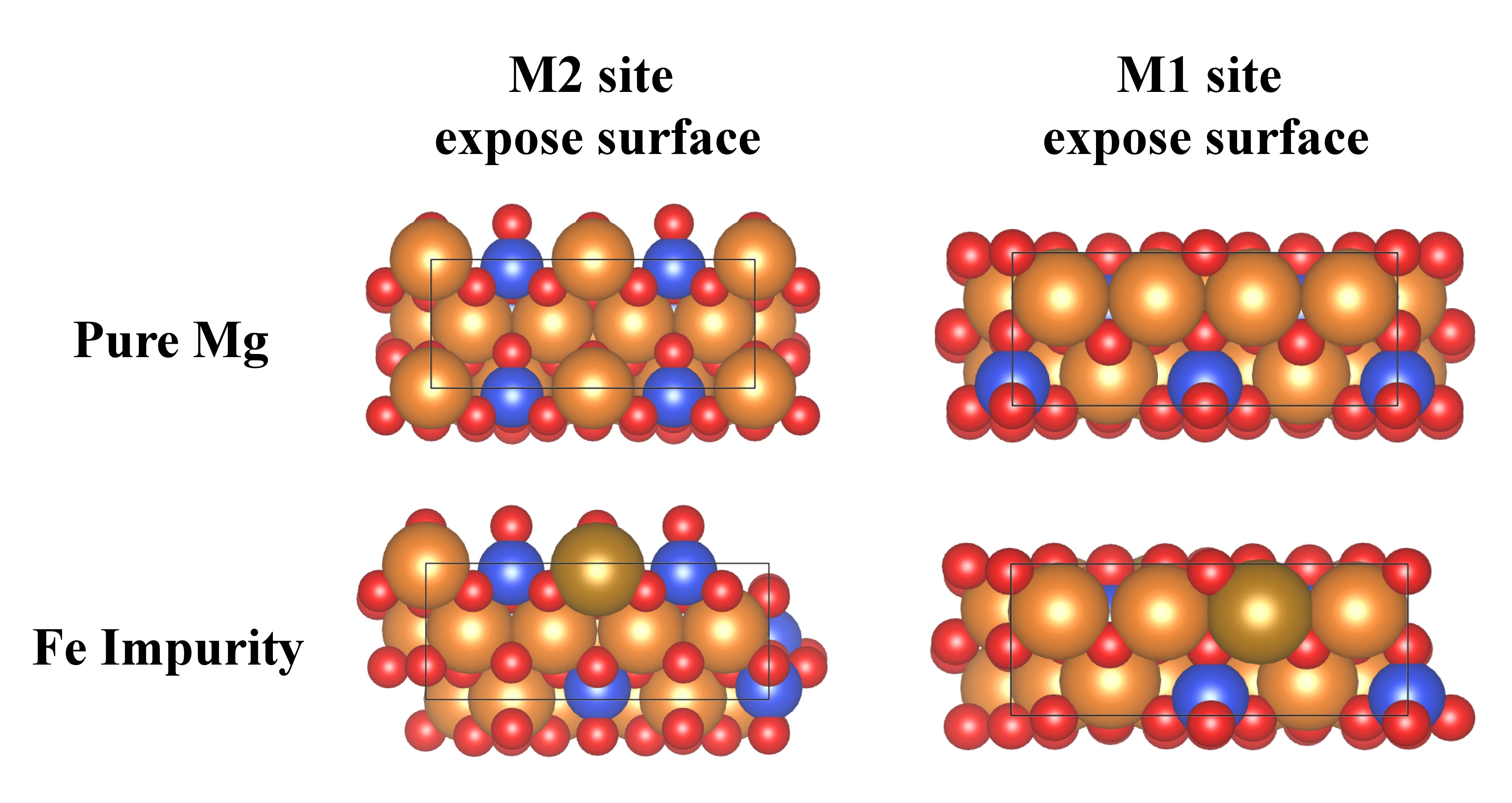}
	\caption{Top view of possible slab structures with/without Fe atom.}
	\label{fig:TopSurface}
\end{figure*}

\end{document}